\documentclass[prl,twocolumn,showpacs,preprintnumbers,amsmath,amssymb]{revtex4-1}
\usepackage{graphicx}
\usepackage{epstopdf}
\usepackage{color}
\usepackage{bbold}
\begin{document}
\title{Pseudo-gap in tunneling spectra as a signature of inhomogeneous superconductivity in oxide interfaces} 
\author{D. Bucheli$^1$, S. Caprara$^{1,2}$, and M. Grilli$^{1,2}$} 
\affiliation{$^1$Dipartimento di Fisica, Universit\`a di Roma Sapienza, piazzale Aldo Moro 5, 
I-00185 Roma, Italy\\
$^2$ISC-CNR and Consorzio Nazionale Interuniversitario per le Scienze Fisiche della 
Materia, Unit\`a di Roma Sapienza, Italy\\
}

\begin{abstract} 
We present a theory for the pseudo-gap state recently observed at the LaAlO$_3$/SrTiO$_3$ 
interface, based on superconducting islands embedded in a metallic background. Superconductivity within each 
island is BCS-like, and the local critical temperatures are randomly distributed, some of them necessarily 
exceeding the critical temperature for global percolation to the zero resistance state. Consequently, tunneling 
spectra display a suppression of the density of states and coherence peaks already well above the percolative 
transition. This model of inhomogeneous superconductivity accounts well for the experimental tunneling spectra. 
The temperature dependence of the spectra suggests that a sizable fraction of the metallic background becomes 
superconducting by proximity effect when the temperature is lowered.
\end{abstract} 
\date{\today} 
\pacs{74.81.-g, 74.78.Fk, 74.55.+v, 74.20.De} 
\maketitle 
 
The observation of a two-dimensional (2D) metallic state at the interface of two insulating oxides 
\cite{ohtomo,Mannhart:2008uj,Mannhart:2010ha,Hwang:2012nm}, and the subsequent demonstration of its 
gate-tunable metal-to-superconductor transition \cite{reyren,triscone,espci1,espci2}, have attracted much 
attention in the last decade. Numerous experiments indicate that the 2D electron gas (EG) is inhomogeneous: 
Transport measurements reveal a large width of the superconducting (SC) transition, suggesting charge 
inhomogeneity \cite{CGBC,BCCG,caprara,SPIN}, and, at lower carrier density, a saturation to a plateau with finite 
resistance, which is a clear signature of the percolating character of the metal-to-superconductor transition. 
Magnetometry \cite{ariando,luli,bert,metha1,metha2,bert2012}, tunneling \cite{ristic}, and piezo-force 
spectroscopy \cite{feng_bi} experiments report submicrometric inhomogeneities. It seems likely that inhomogeneities 
at nanometric scales \cite{espciNM} coexist with structural inhomogeneities at micrometric scales \cite{Kalisky}.
Into this picture arguably enter the recent superconductor-insulator-metal tunnel spectroscopy measurements 
\cite{Richter}, that detect a state with finite resistance, but SC-like density of states (DOS). The experiments 
are carried out in LaAlO$_3$/SrTiO$_3$ (LAO/STO) interfaces, by depositing a metallic Au electrode on the insulating 
LAO layer, and then driving a tunnel current $I$ (by means of a bias voltage $V$) between the electrode and the 2DEG. 
The size of the electrode measures several hundreds $\mu$m (orders of magnitude larger than the inhomogeneities 
\cite{espciNM}). The carrier density of the 2DEG is tuned by means of a back-gating voltage $V_G$ across the STO 
slab. At very low temperature, $T=30$\,mK, the measurements reveal a gap in the DOS at the Fermi energy ($E_F$) 
over the whole range $V_G\in[-300,300]$\,V, accompanied by more or less pronounced coherence peaks above the gap, 
signaling SC coherence and pairing as the origin of DOS suppression. In the carrier depleted regime ($V_G<0$), the 
suppression occurs even when global superconductivity is absent down to the lowest accessible temperatures, 
thereby highlighting the inhomogeneous character 
of the state formed by regions with SC pairing embedded in the non-SC matrix; the DOS at $E_F$ is diminished 
and the coherence peaks are broadened, suppressed, and shifted to slightly higher energies upon 
decreasing $V_G$. At very low carrier concentration, $V_G\approx -300$\,V, the coherence peaks have practically 
vanished, but a substantial gap is still present as a signature of (incoherent) pairing \cite{nota_pseudogap}. 
The temperature dependence of the DOS indicates a peculiar pseudo-gap behavior with a suppression persisting 
above the critical temperature at which the overall resistance vanishes (if any). At higher 
carrier density, $V_G=200$\,V, the 2DEG displays a SC gap and coherence peaks that decrease 
with temperature and vanish around $300$\,mK (which agrees with the critical temperature of bulk STO reported in 
Ref. \cite{koonce}). The dependence grows more complex at $V_G=0$, and even more so at negative $V_G$, where the gap 
and the coherence peaks vanish, respectively, at $T\approx 400$\,mK and $T\approx 600$\,mK, remarkably, 
well above the respective global $T_c$.
In the present Letter, we aim to explain these observations in a simple and coherent framework, based on the
inhomogeneous character of the 2DEG. Since the measurements are taken over several hundreds $\mu$m, we propose 
that the measured pseudo-gap is the result of an average over SC regions (with a DOS described by standard 
BCS theory) and metallic regions with constant DOS $N_0$. 

The equation relating the differential conductance d$I/$d$V$ to the DOS $\rho$ reads:
\begin{equation}\label{dIdV}
\frac{\text{d}I}{\text{d}V}\bigl(V\bigr)=G_0+G_1\int_{-\infty}^{\infty} \rho(E)\,\frac{\partial f(E+eV)}{\partial V}
\,\text{d}E,
\end{equation}
where $f(E)=(1+\text{e}^{E/T})^{-1}$ is the Fermi function, $e$ is the electron charge, the constant $G_0>0$ 
customarily accounts for measurement errors such as leakage currents, and $G_1$ is a dimensional constant. 
In our model, the 2DEG consists of a metallic background embedding islands in which SC pairing occurs below a 
local critical temperature $T_c$, randomly distributed with a probability distribution $P(T_c)$. We tested 
various distributions \cite{SM}, but the resulting DOS do not differ significantly, provided that the mean and width of 
the distributions are comparable. This holds especially at high $T$, where the fine details of the distribution 
are smeared by thermal broadening. In the following, for the sake of definiteness, we take $P(T_c)$ as a 
Gaussian distribution with mean $\overline{T}_c$ and variance $\sigma$, and denote by $w\in[0,1]$ the fraction 
of the sample occupied by the islands. The DOS of the 2DEG can be subdivided into three contributions:
\begin{eqnarray}\nonumber
\rho(E)&=&(1-w)N_0+wN_0\int_{-\infty}^{T}\text{d}T_c\,P(T_c)\\ \label{rho}
&+&w\int_{T}^{\infty}\text{d}T_c\,P(T_c)\,\rho_{\Delta(T_c,T)}(E).
\end{eqnarray}
The first two terms correspond, respectively, to the metallic background and to islands where pairing has 
not taken place yet,  and give an additional constant contribution to the differential conductivity. As shown below,
this contribution fully accounts for the background in the tunneling spectra, allowing us to discard $G_0$ in Eq. 
(\ref{dIdV}). If the $T_c$ distribution is so broad to be non-zero also for $T_c<0$, the second 
term stays finite even at $T=0$. In this case, the total fraction of the system that can display pairing down 
to $T=0$, $w_{pair}$, is smaller than $w$. The third term corresponds to islands which have a finite pairing 
gap $\Delta$. We take the DOS of these islands to be of the form:
\begin{equation}\nonumber
\rho_{\Delta}(E)=N_0\biggl[(1-x)\,\frac{|E|}{\sqrt{E^2-\Delta^2}} + x\Biggr]\Theta\bigl(|E|-\Delta\bigr),
\end{equation}
$\Theta(E)$ being the Heaviside function. The first term is the standard BCS expression, and describes 
coherent pairing occurring in a $(1-x)$ fraction of the whole gapped part. The second term describes islands 
which have a finite gap, but are too small to exhibit well-established phase coherence. Thus, we define a 
coherently paired SC fraction $w_{coh}=(1-x)\,w_{pair}$ and an incoherently paired fraction $w_{inc}=x\,w_{pair}$. 
The introduction of the latter term is motivated by the experimental conductance curves which, in the regime 
$V_G\ll 0$, exhibit well-formed gaps, but practically no coherence peaks (see Fig.\,3a in Ref.\,\cite{Richter}). 
For the gap we borrow the approximated BCS expression:
\begin{equation}\label{bcsgap}
\Delta(T_c,T)=1.76\,T_c\,\text{Tanh}\biggl(\frac{\pi}{1.76}\sqrt{\frac{T_c}{T}-1}\,\biggr).
\end{equation}
The value of $N_0$ is fixed by the high-bias data \cite{notaN0}; $\overline{T}_c$, $\sigma$, $w$, and $x$ are 
adjusted at each $V_G$ to yield the best fit.

\begin{figure}[h]
\centering
\includegraphics[scale=0.18]{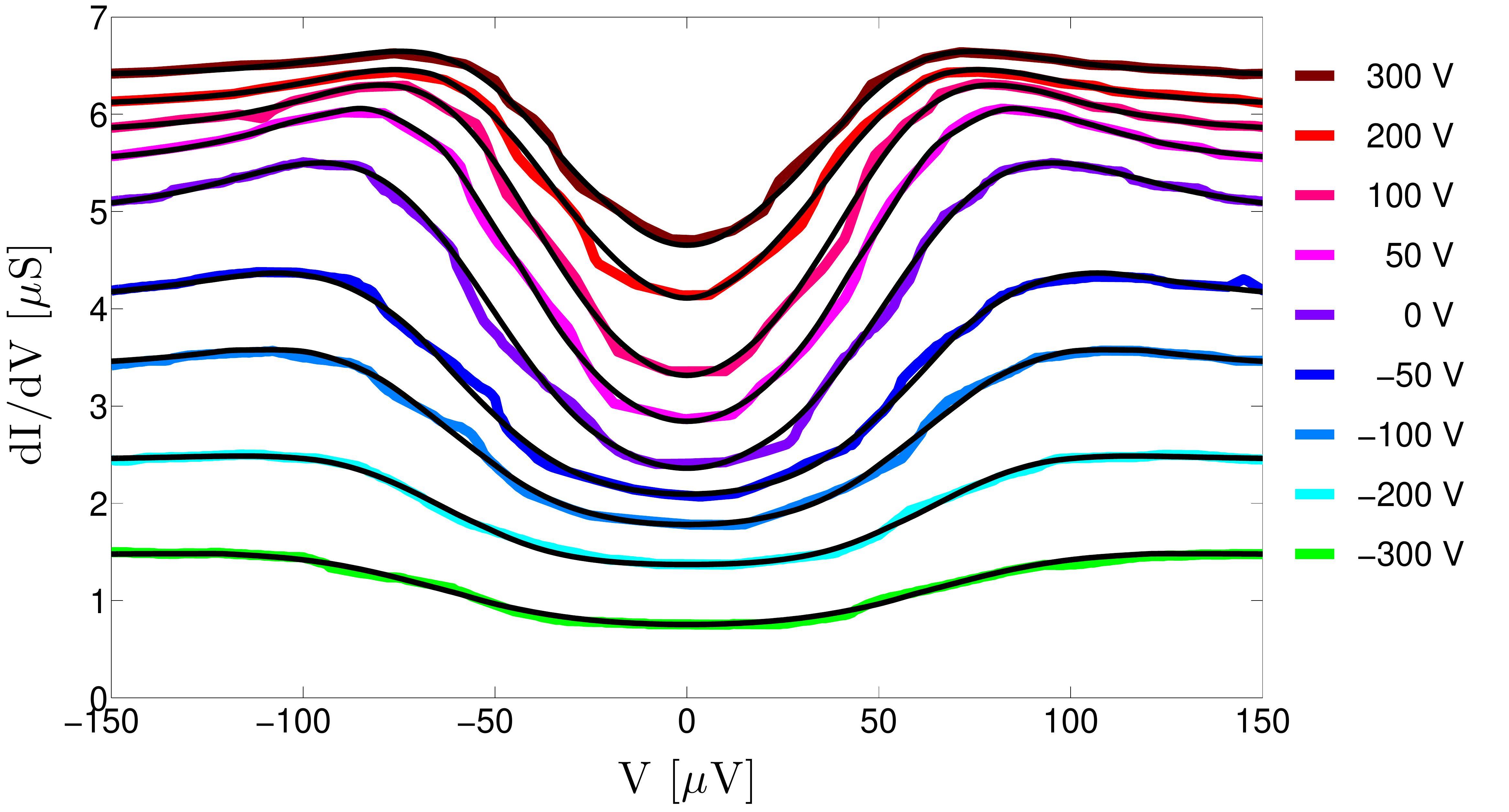}
\caption{
Fits (black curves) of the experimental tunneling data of Ref.\,\cite{Richter} (colored curves) with a 
Gaussian distribution of $T_c$. The fitting parameters are reported in 
Fig.\,\ref{Fig3a_3w_FittingParameters}. The curves at positive (negative) gating have been shifted 
vertically by $+0.3\,\mu$S ($-0.3\,\mu$S) for a better view.}
\label{Fig3a_3w_FinalFigure}
\end{figure}

{\it --- Low temperature behavior ---}We calculate
the best fits of the tunneling spectra at $T=30$\,mK. Fig.\,\ref{Fig3a_3w_FinalFigure} shows that the fits
are in remarkable agreement with the experimental data over the whole range of gating. Our model reproduces 
all the characteristics of the low-$T$ tunneling spectra, from the suppression of the gap minimum and 
the occurrence of coherence peaks in the higher carrier density regime, to the broadening and eventual suppression 
of the coherence peaks at lower carrier density.

\begin{figure}[h]
\centering
\includegraphics[scale=0.24]{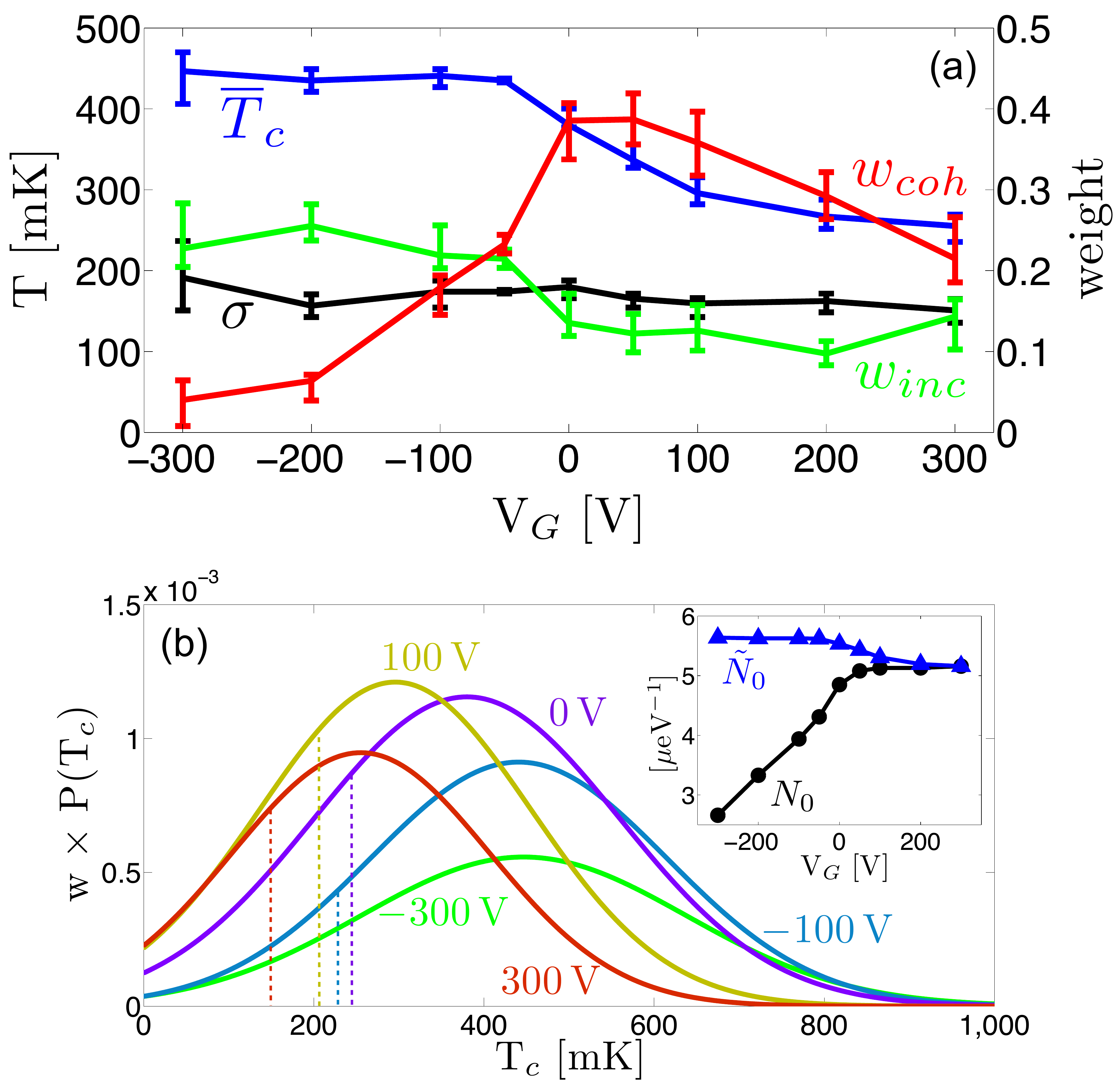}
\caption{
(a)
Parameters of the Gaussian fit. Left $y$-axis: mean critical temperature $\overline{T}_c$ (blue) and variance 
$\sigma$ (black). Right $y$-axis: weight of coherent ($w_{coh}$, red) and incoherent ($w_{inc}$, green) islands.
The error bars define the interval for which the fit differs less than $10$ percent from the best fit 
shown in Fig.\,\ref{Fig3a_3w_FinalFigure}.
(b)
Gaussian distribution of $T_c$ (rescaled with the weight $w$ of SC islands) at various gating. The 
vertical dashed lines indicate the percolative transition to the zero resistance state, if any.
\emph{Inset:} Evolution of the DOS with gating. The DOS of the metal is denoted by $N_0$ (black circles);
$\tilde{N}_0$ (blue triangles) corresponds to the theoretical DOS of the SC islands, calculated from the BCS expression 
$T_c=w_D\,\text{exp}[-1/(g\tilde{N}_0)]$, with $T_c=\overline{T}_c$, $w_D=0.02$\,eV \cite{caprara}, and $g$ taken such 
that $N_0$ and $\tilde{N}_0$ coincide at $V_G=300$\,V.}
\label{Fig3a_3w_FittingParameters}
\end{figure}
The fitting parameters $\overline{T}_c$, $\sigma$, $w$ and $x$, are reported in Fig.\,\ref{Fig3a_3w_FittingParameters}. 
First of all, we notice that increasing $V_G$ (i.e., the carrier density) steadily reduces the average 
local pairing temperature $\overline{T}_c$. This behavior is at odds with the idea that disorder alone rules the 
SC physics in these interfaces: Increasing the carrier density (i.e., $E_F$) effectively reduces the disorder 
parameter $(\tau E_F)^{-1}$ ($\tau$ being the scattering time) and, consequently, $\overline{T}_c$ should increase 
\cite{finkelstein}. Therefore, the decrease of $\overline{T}_c$ with $V_G$ must necessarily be attributed 
to a decrease of the effective pairing potential $\lambda\equiv g\tilde{N}_0$, where $\tilde{N}_0$ denotes the DOS 
in the islands with pairing and $g$ is the BCS coupling. Due to the inhomogeneous nature of the system, $\tilde{N}_0$ 
differs from the DOS of the metallic background, $N_0$. While the latter steadily increases with gating, the former 
must (slightly) decrease to yield the observed decreasing $\overline{T}_c(V_G)$. Moreover, the average 
pairing temperature at, e.g., $V_G=-300$V, $\overline{T}_c\approx 450$\,mK (with some rare regions having 
local $T_c\approx\overline{T}_c+2\sigma\approx 700$\,mK) is somewhat larger than the SC critical temperature 
usually reported for bulk doped STO \cite{koonce,schooley,behnia}.
Larger $T_c$s suggest that pairing is more effective in this low-density 2DEG than in the bulk system. Although we adopt 
here a purely phenomenological approach and do not attempt any detailed microscopic interpretation, it is reasonable to 
attribute the stronger pairing to DOS effects, like a peak in $\tilde{N}_0$ \cite{notaDOS}, effectively enhancing pairing. 
To estimate $\tilde{N}_0$ we proceed as follows. We assume that a BCS-like pairing occurs locally and that 
$\tilde{N}_0=N_0$ at $V_G\approx 300$\,V. Thus, we fix $g=0.028\,\mu$eV to produce the corresponding average critical 
temperature $\overline{T}_c\approx 210$\,mK. Keeping $g$ fixed, we then calculate $\tilde{N}_0$ such that 
$\lambda \equiv g \tilde{N_0}$ yields the fitted $\overline{T}_c$s at all $V_G$s. The resulting $\tilde{N}_0(V_G)$ is 
reported in the inset of Fig.\,\ref{Fig3a_3w_FittingParameters}(b) together with the DOS $N_0$ of the metallic background.

Concerning SC coherence, we find that in the regime $V_G<0$ the islands with paired electrons contain a 
large fraction $w_{inc}$ of regions {\it without} phase coherence. This incoherent fraction can be related to 
paired regions of size $L$ smaller than the SC coherence length $\xi_0\sim 100$\,nm \cite{behnia}, 
and decreases when the carrier density 
is increased, but is never less than $10$ percent. On the other hand, the coherent SC fraction, $w_{coh}$, increases 
and reaches a maximum of $0.4$ between $V_G=0$ and $V_G=100$\,V. Surprisingly, further increasing the 
carrier density at $V_G\ge 200$\,V leads to a decrease of the coherent fraction (and, possibly, a small increase 
of the incoherent but gapped fraction). The fact that $w_{coh}$ stays below 0.5 despite a complete percolation of 
the system, which displays vanishing overall resistance for $V_G>-150$\,V, strongly suggests that the 
distribution of coherent SC regions is spatially correlated, as already found in Refs. 
\cite{BCCG,caprara}. The overall paired fraction, $w_{pair}=w_{coh}+w_{inc}$, increases 
with $V_G$, when $V_G<0$, in accordance with the idea that the SC fraction increases with the carrier density, 
but saturates above $V_G\approx 0$\,V, and even slightly decreases at large positive $V_G$. 
The explanation for this unexpected behavior may rest upon the mechanism leading to the inhomogeneous state.

Noticeably, the width $\sigma$ of the $T_c$ distribution stays nearly constant [see 
Fig.\,\ref{Fig3a_3w_FittingParameters}(b)], showing that this distribution is an intrinsic structural property 
of the sample, likely related to the local random distribution of impurities and defects, which rules the 
(local) transition temperature, as commonly occurs in homogeneously disordered superconductors \cite{finkelstein}. 

\begin{figure}[h]
\centering
\includegraphics[scale=0.23]{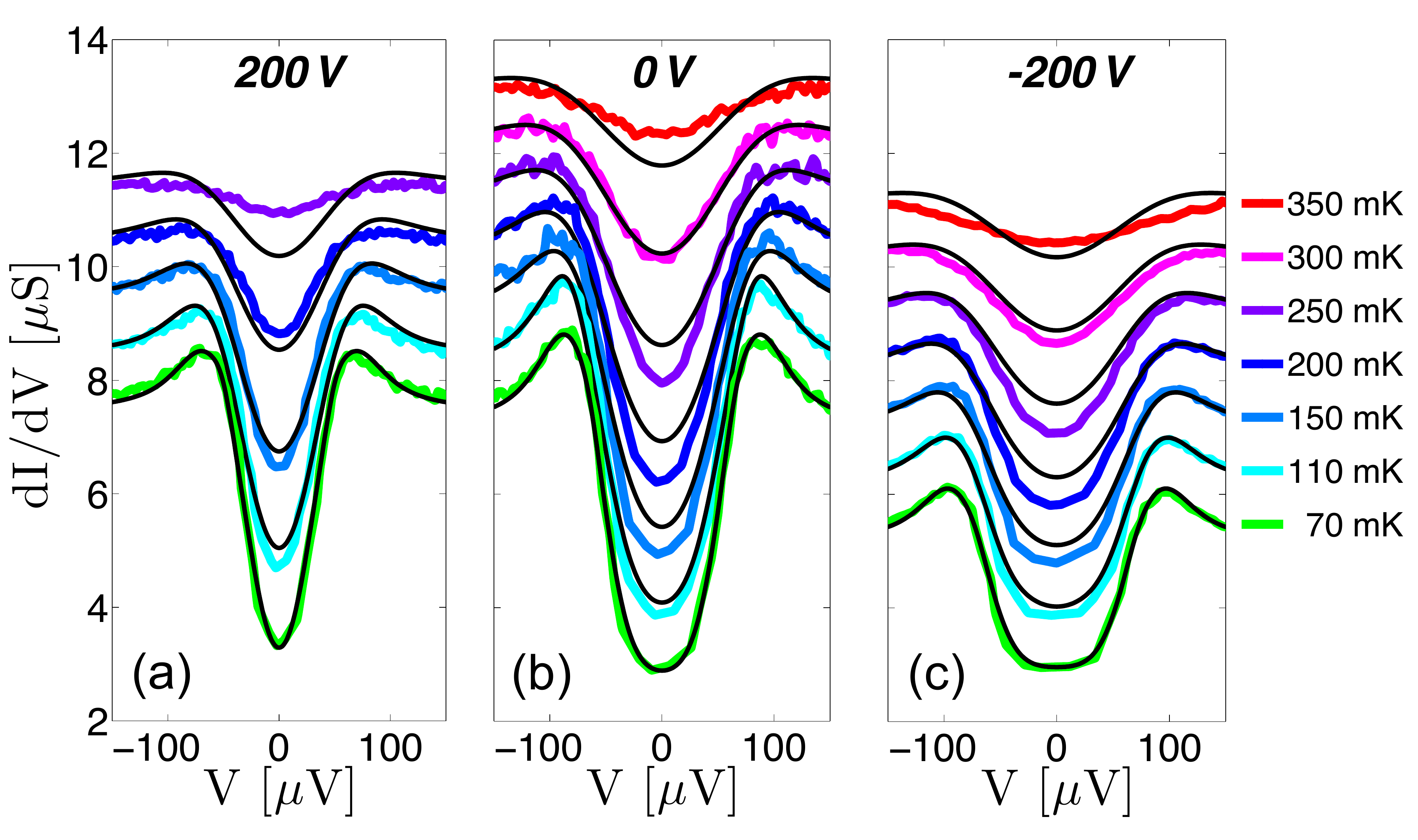}
\caption{
Fits (black curves) of the experimental data of Ref.\,\cite{Richter} (colored curves) with a Gaussian 
distribution for $V_G=200$\,V (a), $V_G=0$\,V (b), $V_G=-200$\,V (c). The curves at $T>70$\,mK have each 
been shifted upwards by $1$\,$\mu$S with respect to the preceding curve for a better view. 
Fitting parameters: (a) $\overline{T}_c=235$\,mK, 
$\sigma=150$\,mK, $w=0.67$, and $x=0.1$; (b) $\overline{T}_c=400$\,mK, $\sigma=135$\,mK, $w=0.6$, and 
$x=0.05$; (c) $\overline{T}_c=465$\,mK, $\sigma=132$\,mK, $w=0.41$, and $x=0.05$.}
\label{Fig3abc_Tindep}
\end{figure}

{\it --- Temperature dependence of the spectra ---}
Next, we turn to the evolution of the spectra as a function of temperature, at fixed $V_G$. We point out that the 
temperature dependence was measured in different experimental runs, months after the low-$T$ spectra in 
Fig.\,\ref{Fig3a_3w_FinalFigure} were measured \cite{Richter}, so that some aging of the sample cannot be excluded. 
As it can be seen in Fig.\,\ref{Fig3abc_Tindep}, our model captures the overall effect of temperature, although 
the fits are less convincing than those in Fig.\,\ref{Fig3a_3w_FinalFigure}. 

\begin{figure}[h]
\centering
\includegraphics[scale=0.23]{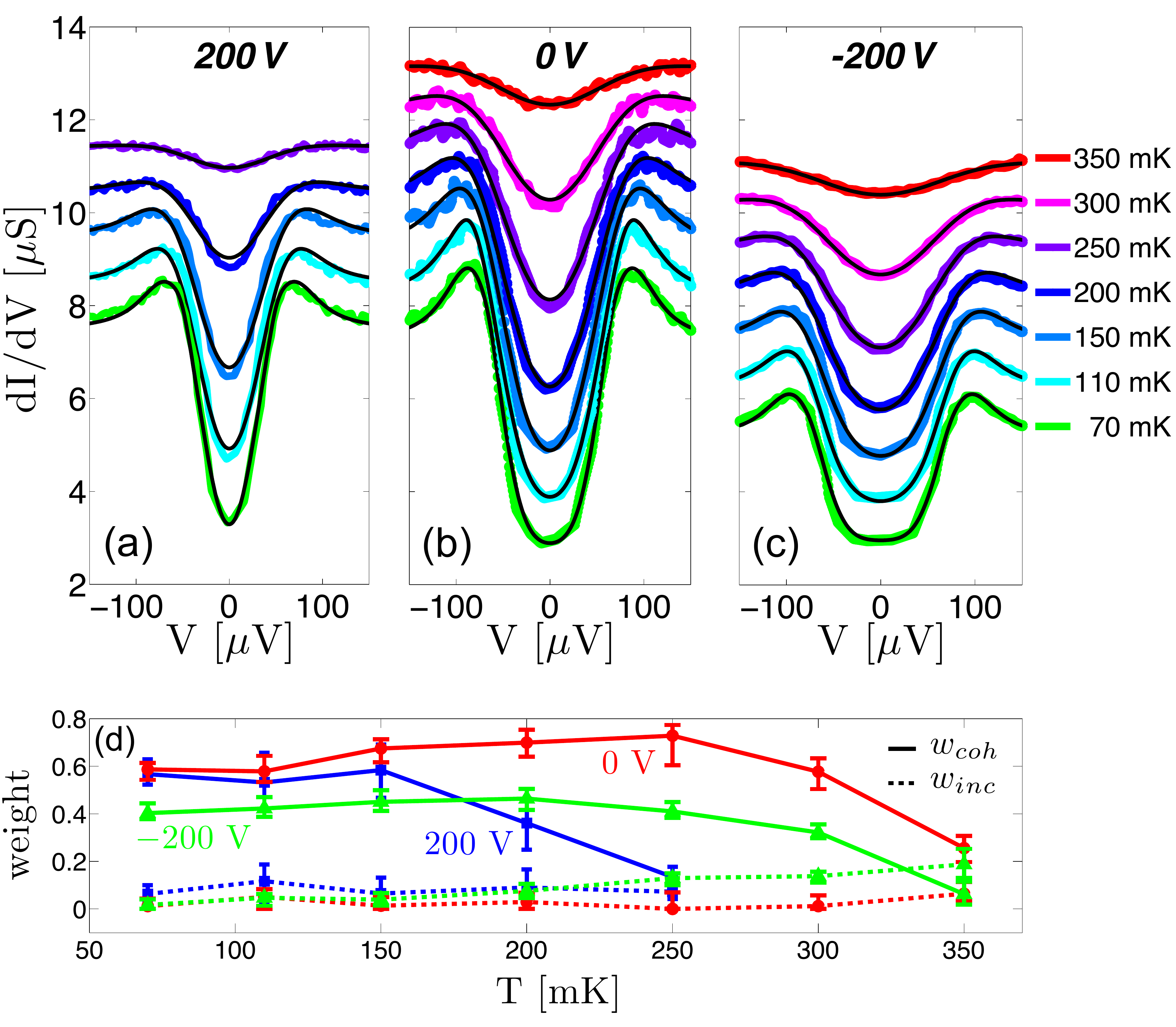}
\caption{ Fits (black curves) of the experimental data of Ref.\,\cite{Richter} (colored curves) 
with a Gaussian distribution for $V_G=200$\,V (a), $V_G=0$\,V (b), $V_G=-200$\,V (c).
The curves at $T>70$\,mK have each been shifted upwards by $1$\,$\mu$S with respect to the preceding curve for 
a better view. $\overline{T}_c$ and $\sigma$ are the same as in Fig.\,\ref{Fig3abc_Tindep}. (d) 
Weight of coherent ($w_{coh}$, solid lines) and incoherent ($w_{inc}$, dotted lines) islands as a function 
of temperature.}
\label{Fig3abc_NEW}
\end{figure}

This is certainly due to the very strong constraint of $T$-\emph{independent} fitting parameters,
which is not strictly supported by the data. For instance, at $V_G=200$\,V, the depth and width of the conductance 
curve at low $T$ suggests a substantial gap $\Delta\approx 60$\,$\mu$eV, which, according to Eq. 
(\ref{bcsgap}), corresponds to a critical temperature $T_c\approx 340$\,mK. However, already at
$T\approx 250$\,mK, the experimental curves display only a minor DOS suppression, while, according to BCS theory, the 
regions with sizable low-$T$ gaps (i.e., with large $T_c$s) are still SC and yield too large a DOS suppression. 
This calls for an improvement of the model to account for substantial low-$T$ gaps, which however seem to vanish 
at temperatures lower than expected by standard BCS theory. A natural possibility is that the overall paired fraction 
is not fully established at high $T$ by structural properties alone. It is instead conceivable that, 
upon lowering the temperature, the proliferation of SC regions influences by proximity effect the metallic background, 
thereby causing an increase in $w_{pair}$. To investigate this possibility, we relaxed the constraint of a 
$T$-independent paired fraction, allowing both $w_{coh}$ and $w_{inc}$ (i.e., $w$ and $x$) to be adjusted at 
each temperature, while the parameters of the $T_c$ distribution are kept fixed. The resulting fits, 
reported in Fig.\,\ref{Fig3abc_NEW}, are now in very good agreement with the experimental data.
The improvement is obtained by letting $w_{pair}$ to augment up to $40$ percent when $T$ is lowered, 
mostly due to an increase of $w_{coh}$. Below $250$\,mK, for $V_G=0$\,V and $V_G=-200$\,V, or below $150$\,mK, 
for $V_G=200$\,V, the paired fractions saturate to nearly constant values.

We tested alternative ways to improve the fits of Fig.\,\ref{Fig3abc_Tindep}. First, we considered deviations 
from the BCS temperature dependence of the gap in Eq. (\ref{bcsgap}), by introducing a temperature dependent 
prefactor, which should be larger than 1.76 in the case of strong-coupling pairing \cite{SM}. 
However, despite a generic improvement and a smaller width of the distributions, we were unable to reach the quality 
of the fits in Fig.\,\ref{Fig3abc_NEW}. Moreover, the occurrence of strong-coupling effects seems rather unlikely, 
because the estimated values of the effective pairing potential, $\lambda\sim 0.07-0.14$, are rather small. Finally, 
we considered a $T$-dependent inverse lifetime for the Cooper pairs, $\Gamma(T)$, according to the so-called 
Dynes formula, used in the fits of Ref.\,\cite{Richter}. Also in this case, and quite expectedly, the broadening 
due to $\Gamma(T)$ allows to take narrower $T_c$ distributions ($\sigma$ is even reduced by fifty 
percent). Rather good fits are obtained \cite{SM}, but again not as good as in 
Fig.\,\ref{Fig3abc_NEW}, with values of $\Gamma(T)$ much smaller than those reported in Ref.\,\cite{Richter}, 
and always smaller than $20$\,$\mu$eV. 

{\it --- Conclusions ---}
Our work demonstrates that the recent observation of a pseudo-gap in LAO/STO is explained by the well established 
inhomogeneous character of the 2DEG in this system. Specifically, a metallic background embeds regions which display
pairing below randomly distributed $T_c$s. The features of the $T_c$ distribution are structurally determined 
and therefore do not depend on temperature. Inside most of these regions, standard BCS coherence is established,
while a smaller fraction only displays incoherent pairing, likely associated with the size $L$ of these islands 
being smaller than the SC coherence length $\xi_0$. This indicates that inhomogeneities in these 2DEG also 
occur on length scales smaller than 100\,nm. Attempts to fit the temperature dependence of the spectra 
also lead to the conclusion that some (30-40 percent) of the paired regions at low temperature are likely formed 
by proximity effect in the metallic matrix and were not ``foreseen" at higher temperature. The importance of 
proximity for establishing superconductivity in the quantum critical regime at low carrier density in 
LaTiO$_3$/STO was also assessed in Ref. \cite{espciNM}. The introduction of other 
physical effects (deviations from the BCS temperature dependence of the gap, pair-breaking effects, ...) may 
improve the fits and lead to a reduction of the temperature dependence of $w_{pair}$, but the scenario reported 
here is rather robust. In conclusion, the very recognition of the inhomogeneous character of these systems 
naturally leads to a direct interpretation of the pseudo-gap effects observed in tunneling experiments without 
exotic non-BCS physical ingredients. 

\begin{acknowledgements} 
We acknowledge insightful discussions with L. Benfatto, N. Bergeal, C. Castellani, J. Lesueur, and G. Seibold,
and financial support from the Progetti AWARDS of the Sapienza University, Project n. C26H13KZS9.
We warmly thank H. Boschker for providing us with the data of the experiments in Ref.\,\cite{Richter}. 
\end{acknowledgements} 
 
\eject
\newpage
\begin{appendix}
\begin{widetext}

\begin{center}
{\bf SUPPLEMENTARY MATERIAL}
\end{center}

\vspace{1 truecm}
\section{Distribution of critical temperatures at low temperature}
In the main text of our Letter we showed that the low temperature behavior of the tunneling spectra can be perfectly fitted with a Gaussian distribution of critical temperatures.
In order to illustrate to what extent the fits depend on the distribution, 
we present the outcomes with three additional distributions:
\begin{eqnarray*}
\text{Box: }P(T_c)&=&\frac{1}{T_c^{max}-T_c^{min}}\,\Theta(T_c^{max}-T_c)\,\Theta(T_c-T_c^{min}),\nonumber\\
\text{Lorentzian: }P(T_c)&=&\frac{\gamma}{\pi\,[\gamma^2+(\overline{T}_c-T_c)^2]},\nonumber\\ 
\text{Skewed Lorentzian: }P(T_c)&=&\frac{1}{\frac{1}{2}\text{Log}\left[ 1+\left(\frac{T_c^{max}}{\gamma}\right)^2\right]}\frac{T_c^{max}-T_c}{\gamma^2+(T_c^{max}-T_c)^2}\Theta(T_c^{max}-T_c)\,\Theta(T_c).\nonumber
\end{eqnarray*}
Below we report the best fits (left panel), the difference between the theoretical and experimental curve (middle panel) and the corresponding $T_c$ distributions (right panel) 
at $T=70$\,mK for $V_G=200$\,V (Fig. \ref{SupplMat_PTc_Vg200}), $0$\,V (Fig. \ref{SupplMat_PTc_Vg0}) and $-200$\,V (Fig. \ref{SupplMat_PTc_Vgm200}).

The Box distribution, being constant in the interval $[T_c^{min},T_c^{max}]$ and zero elsewhere, allows for a precise control over the range of critical temperatures. 
However, this strictness is somewhat at odds with the rather broad coherence peaks of the experimental curves.
Indeed, inspection of the three figures below reveals that the fits (yellow curves) result in too accentuated coherence peaks.
Instinctively, one would think that this flaw is cured by taking a larger incoherent part ($w_{inc}=w\,x$). Although this is the case, larger $w_{inc}$ lead to overall worse fits, because increasing $x$ the fits worsen for $V\gtrsim|100|\,\mu$V.

Fitting the experimental data with a Lorentzian distribution (green curve) gives rather convincing agreement between experiment and theory.
The fits exhibit the interesting feature of $w_{inc}=0$, i.e. all the islands with pairing are coherent. 
This is due to the distribution's very large width (its variance is even divergent) which spreads the coherence peaks over a broad range of energies.
On the down side, this large width corresponds to a (rather unrealistic) situation with small but non-negligible weight up to critical temperatures of $1$\,K (especially for low gating).

The effect of asymmetry is studied considering a skewed Lorentzian distribution (blue curve). The fits shown in Figs. \ref{SupplMat_PTc_Vg200}-\ref{SupplMat_PTc_Vgm200} are less convincing compared to the previous two. 
The main reason for the failure is that in order to have enough weight around the main gap (compare for example with the Box distribution) one needs rather large values of $\gamma$ which then give too much weight to low critical temperatures, i.e.
around $0\,\mu$V. At higher energies, the fits exhibit the same shortcoming of too strong coherence peaks as did the fits with the Box distribution, due to the sharp decrease of $P(T_c)$ at 
$T_c^{max}$.

In conclusion, the Gaussian distribution (red curve) yields the best fits, having in some sense the advantages of both the Box distribution and the Lorentzian.
While the former has equal weight in a certain interval of $T_c$s, leading to good fits at intermediate energies ($\sim50\,\mu$eV), the latter allows for a good description of
the center of the spectra ($\sim0\,\mu$eV) and the coherence peaks ($\sim100\,\mu$eV).
\begin{figure}[!htbp]
\centering
\includegraphics[scale=0.5]{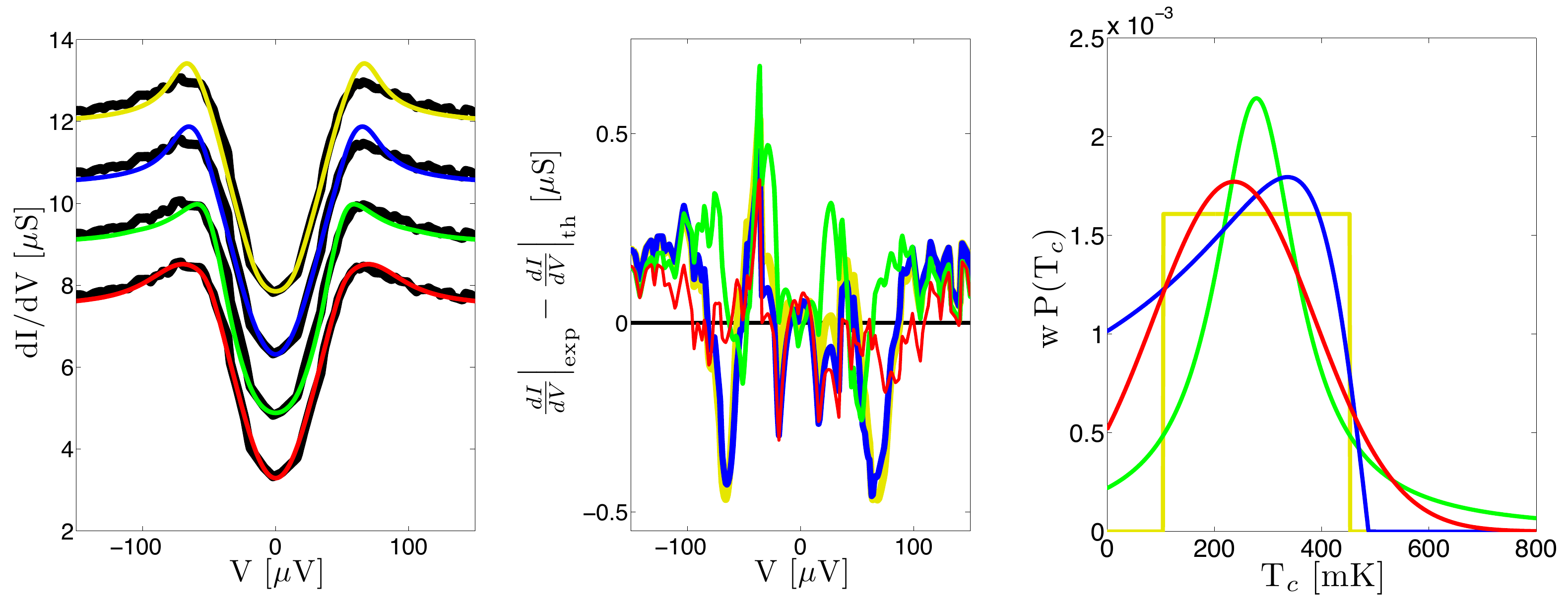}
\caption{
Comparison of $P(T_c)$ at low temperature for $V_G=200$\,V. 
The black curves correspond to the experimental tunneling data of Ref. \cite{Richter} and have been shifted vertically by $1.5\,\mu$S for a better view.
The colored curves correspond (from bottom to top) to
Gaussian ($\overline{T}_c=237$\,mK, $\sigma=151$\,mK, $w=0.67$, $x=0.1$);
Lorentzian ($\overline{T}_c=27.8$\,mK, $\gamma=93$\,mK, $w=0.64$, $x=0$);
Skewed Lorentzian ($T_c^{max}=487$\,mK, $\gamma=151$\,mK, $w=0.66$, $x=0.1$);
Box ($T_c^{min}=104$\,mK, $T_c^{max}=452$\,mK, $w=0.56$, $x=0.15$).
}
\label{SupplMat_PTc_Vg200}
\end{figure}
\begin{figure}[!htbp]
\centering
\includegraphics[scale=0.5]{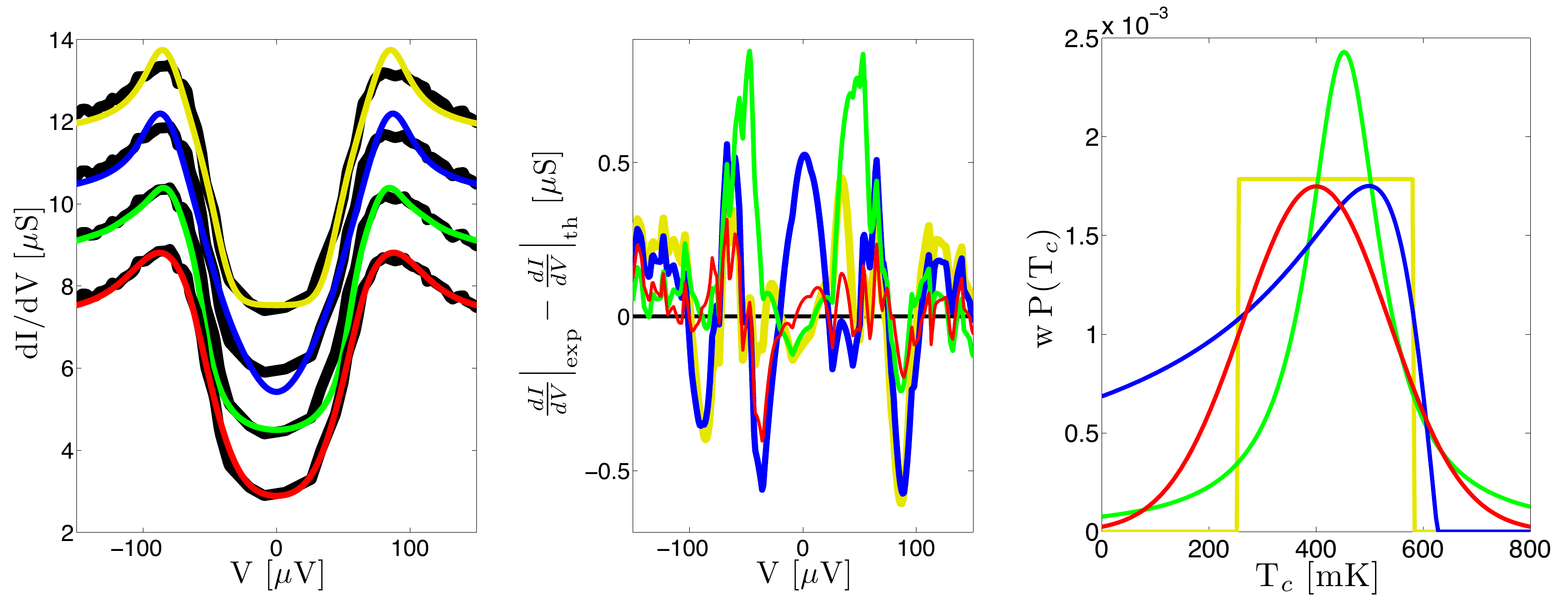}
\caption{
Comparison of $P(T_c)$ at low temperature for $V_G=0$\,V. 
The black curves correspond to the experimental tunneling data of Ref. \cite{Richter} and have been shifted vertically by $1.5\,\mu$S for a better view.
The colored curves correspond (from bottom to top) to
Gaussian ($\overline{T}_c=400$\,mK, $\sigma=137$\,mK, $w=0.6$, $x=0.05$);
Lorentzian ($\overline{T}_c=452$\,mK, $\gamma=81$\,mK, $w=0.62$, $x=0$);
Skewed Lorentzian ($T_c^{max}=626$\,mK, $\gamma=128$\,mK, $w=0.72$, $x=0.1$);
Box ($T_c^{min}=255$\,mK, $T_c^{max}=580$\,mK, $w=0.58$, $x=0.1$).
}
\label{SupplMat_PTc_Vg0}
\end{figure}
\begin{figure}[!htbp]
\centering
\includegraphics[scale=0.5]{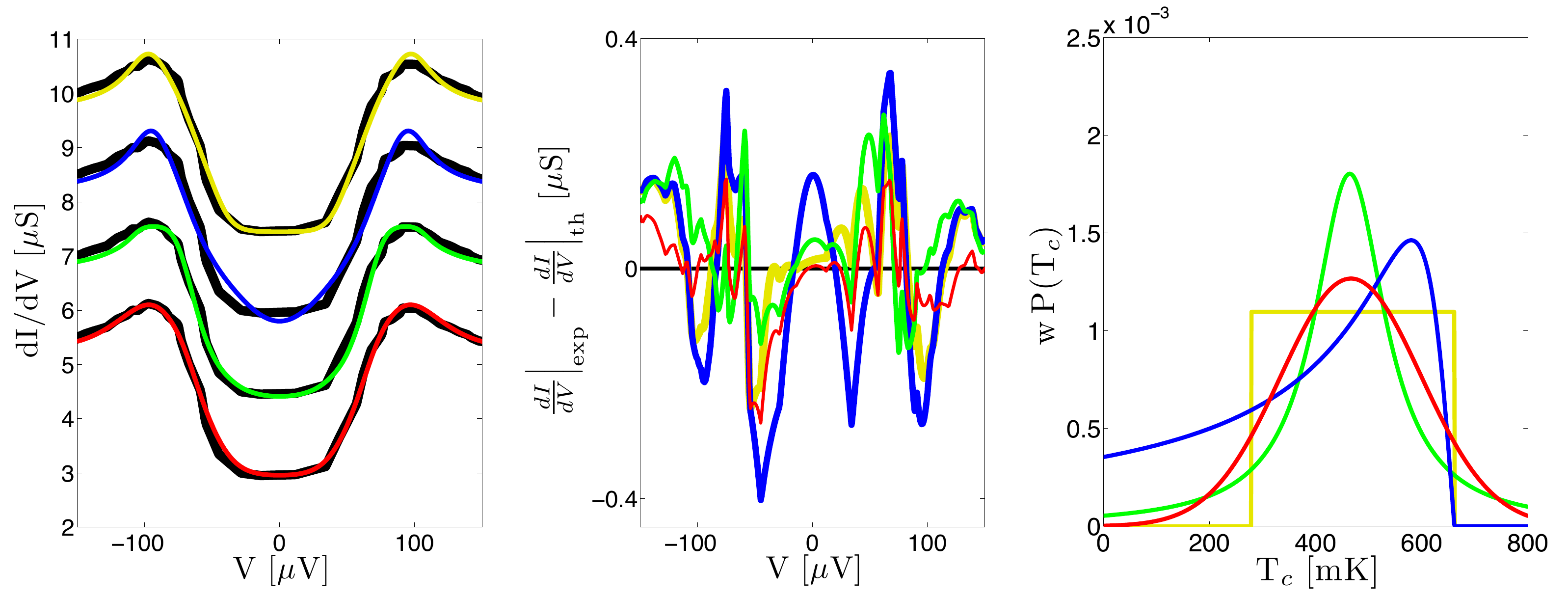}
\caption{
Comparison of $P(T_c)$ at low temperature for $V_G=-200$\,V. 
The black curves correspond to the experimental tunneling data of Ref. \cite{Richter} and have been shifted vertically by $1.5\,\mu$S for a better view.
The colored curves correspond (from bottom to top) to
Gaussian ($\overline{T}_c=466$\,mK, $\sigma=133$\,mK, $w=0.42$, $x=0.05$);
Lorentzian ($\overline{T}_c=580$\,mK, $\gamma=81$\,mK, $w=0.46$, $x=0$);
Skewed Lorentzian ($T_c^{max}=661$\,mK, $\gamma=81$\,mK, $w=0.5$, $x=0.1$);
Box ($T_c^{min}=278$\,mK, $T_c^{max}=661$\,mK, $w=0.42$, $x=0.1$).
}
\label{SupplMat_PTc_Vgm200}
\end{figure}
\clearpage

\section{Behavior in Temperature}
The description of the temperature behavior of the tunneling spectra faces several challenges. 
The main one arises from the connection between large low-temperature gaps and the associated high critical temperatures.
More to the point, the density of states (DOS) suppression at low temperature reveals substantial gaps which, within BCS
theory, correspond to critical temperatures higher 
than the temperature at which the measurements report the closure of the gap. 
Further, the spectra for $V_G=0,-200$\,V slightly \emph{decrease} with temperature before they ultimately increase around $T=150$\,mK.
In both cases, insisting on standard BCS theory and a $T_c$ distribution fixed in temperature, inconsistencies between low- and high-temperature regime are inevitable.
In the light of these experimental features one is bound to conclude that 
additional effects in temperature are present, other than the ones stemming from the Fermi function.
As shown in the main text, the possibility that the coherent ($w_{coh}$) and incoherent ($w_{inc}$) fraction 
may vary with temperature successfully resolves the difficulties raised above.
In the following we investigate three alternative solutions.

The first adjustment consists in taking a distribution of superconducting gaps.
The physical idea behind this approach is simple. The gap of an island may not be exactly uniform (although the island has a single, well-defined critical temperature),
but follows a certain distribution.
For instance, one can assume that the electrons of an island fill up several sub-bands, which, as soon as one of them becomes superconducting, 
become superconducting as well. The strengths of intra- and inter-band coupling being different, this may lead to a distribution of gaps.
Depending on the distribution chosen, this mimics strong coupling effects since a given $T_c$ also yields gaps larger than the BCS prediction.
For the sake of simplicity, we assume that an island has a flat distribution of gaps
$
P(\Delta)=(\Delta^{max}-\Delta^{min})^{-1}\Theta(\Delta^{max}-\Delta)\Theta(\Delta-\Delta^{min})$, 
which leads to the following density of states: 
\begin{eqnarray*}
\rho_{SC}(E,\Delta^{min},\Delta^{max})&=&|E|\int_{-\infty}^{+\infty}d\Delta\frac{P(\Delta)}{\sqrt{E^2-\Delta^2}}
\Theta\bigl(|E| - \Delta \bigr) \nonumber\\
  \nonumber \\
   &=& \frac{|E|}{\Delta^{max}-\Delta^{min}}\left\{
   \begin{array}{ll}
    0 & :\:\: |E| < \Delta^{min}\nonumber\\
    \frac{\pi}{2} - \text{Arcsin}\Bigl(\frac{\Delta^{min}}{|E|}\Bigr) & :\:\: \Delta^{min} < |E| < \Delta^{max}\nonumber\\
    \text{Arcsin}\Bigl(\frac{\Delta^{max}}{|E|}\Bigr) - \text{Arcsin}\Bigl(\frac{\Delta^{min}}{|E|}\Bigr)& :\:\: |E| > \Delta^{max}
   \end{array}
  \right.\nonumber
\end{eqnarray*}
We choose $\Delta^{min}=\Delta^{BCS}$, $\Delta^{max}=(1+X)\Delta^{BCS}$, while $\Delta^{BCS}$ is related to $T_c$ via the standard BCS 
equation. For the sake of definiteness, we present the results for $X=0.4$. The fits with smaller values of $X$ are barely 
distinguishable from the fits with $X=0$, and larger values are hardly justifiable as a slight modification of the BCS 
theory. The $T_c$ distribution is determined by the best fit at $T=70$\,mK. The results obtained with a Gaussian $P(T_c)$ are 
reported in Fig. \ref{SupplMat_P_di_Delta} (orange dashed curves). 
Expectedly, the best fits are obtained with $P(T_c)$ distributions shifted to temperature lower than their $X=0$ counterparts (black solid lines). To some extent, 
this mitigates the discrepancy between the critical temperatures inferred from low-temperature fitting and the actual temperature at which the DOS 
suppression vanishes.
However, at intermediate temperature the fits still require improvement.
\begin{figure}[!htbp]
\centering
\includegraphics[scale=0.45]{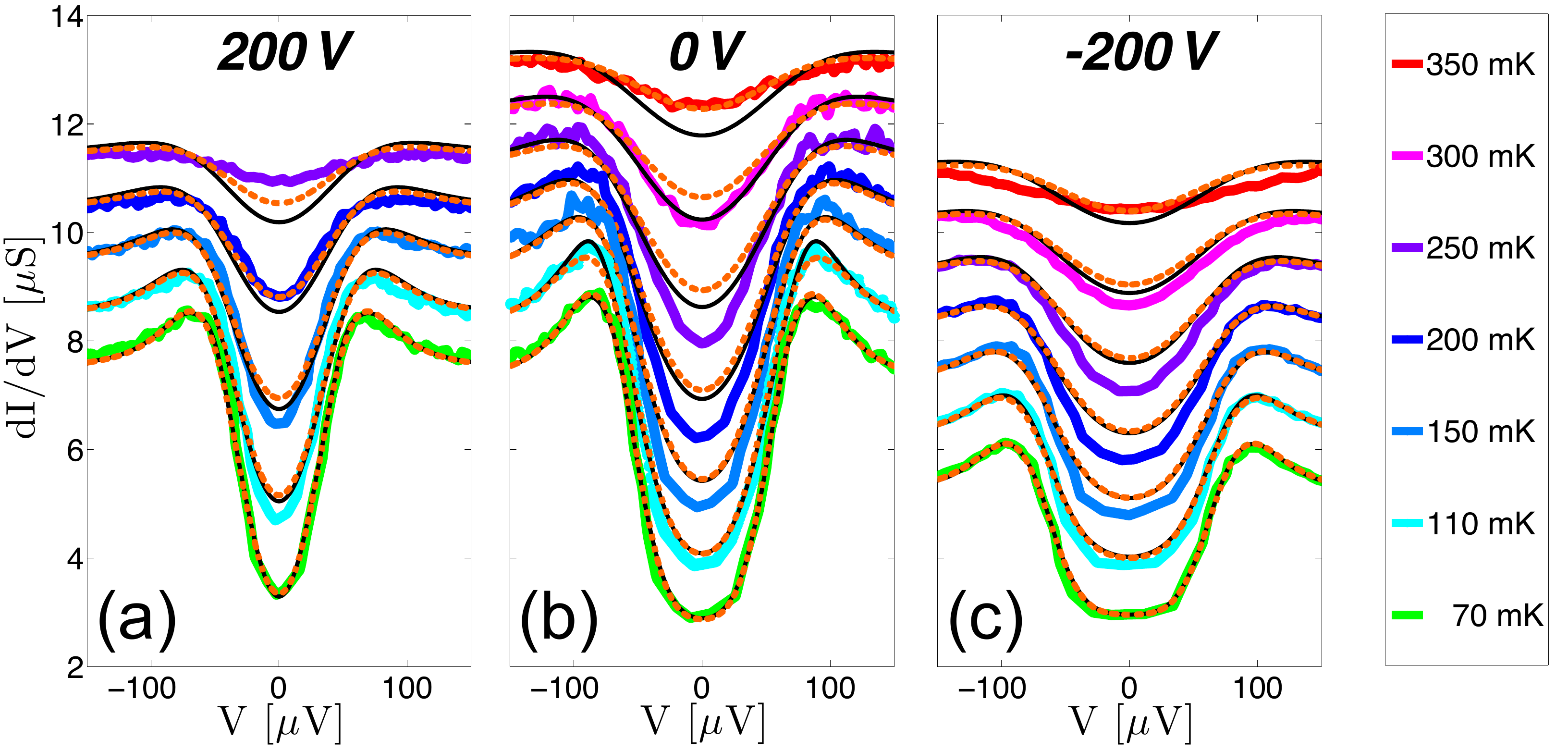}
\caption{Fits of experimental data with $P(\Delta)$ ($X=0.4$) and a Gaussian distribution (orange dashed curves). 
$V_G=200$\,V: $\overline{T}_c=209$\,mK, $\sigma=116$\,mK, $w=0.64$, $x=0.1$;
$V_G=0$\,V: $\overline{T}_c=336$\,mK, $\sigma=104$\,mK, $w=0.6$, $x=0.05$; 
$V_G=-200$\,V: $\overline{T}_c=394$\,mK, $\sigma=104$\,mK, $w=0.42$, $x=0.05$. 
The black curves correspond to the fits without $P(\Delta)$ from Fig. 3 of the main text.}
\label{SupplMat_P_di_Delta}
\end{figure}

The straightforward extension is to take explicitly strong coupling deviations from BCS that may vary in temperature.
To this end we introduce a factor $\alpha(T)$ in the standard BCS expression,
\begin{equation}\label{bcsgap}
\Delta(T_c,T)=1.76\,\alpha(T)\,T_c\,\text{Tanh}\Biggl(\frac{\pi}{1.76}\sqrt{\frac{T_c}{T}-1}\,\Biggr).
\end{equation}
Taking a Gaussian $P(T_c)$ we determine $\overline{T}_c$, $\sigma$ and $w$ (fixed in temperature) such as to minimize the overall difference 
between the theoretical and the experimental curves for all temperatures
while the factor $\alpha(T)$ is let free to take values in the interval $[0.8,1.3]$.
Although values $\alpha(T)<1$ do not correspond to a strong coupling regime, we allowed for this possibility for the sake 
of comprehension. The results given in Fig. \ref{SupplMat_alpha_di_T} show an improvement with respect to $\alpha\equiv 1$. 
\begin{figure}[!htbp]
\centering
\includegraphics[scale=0.42]{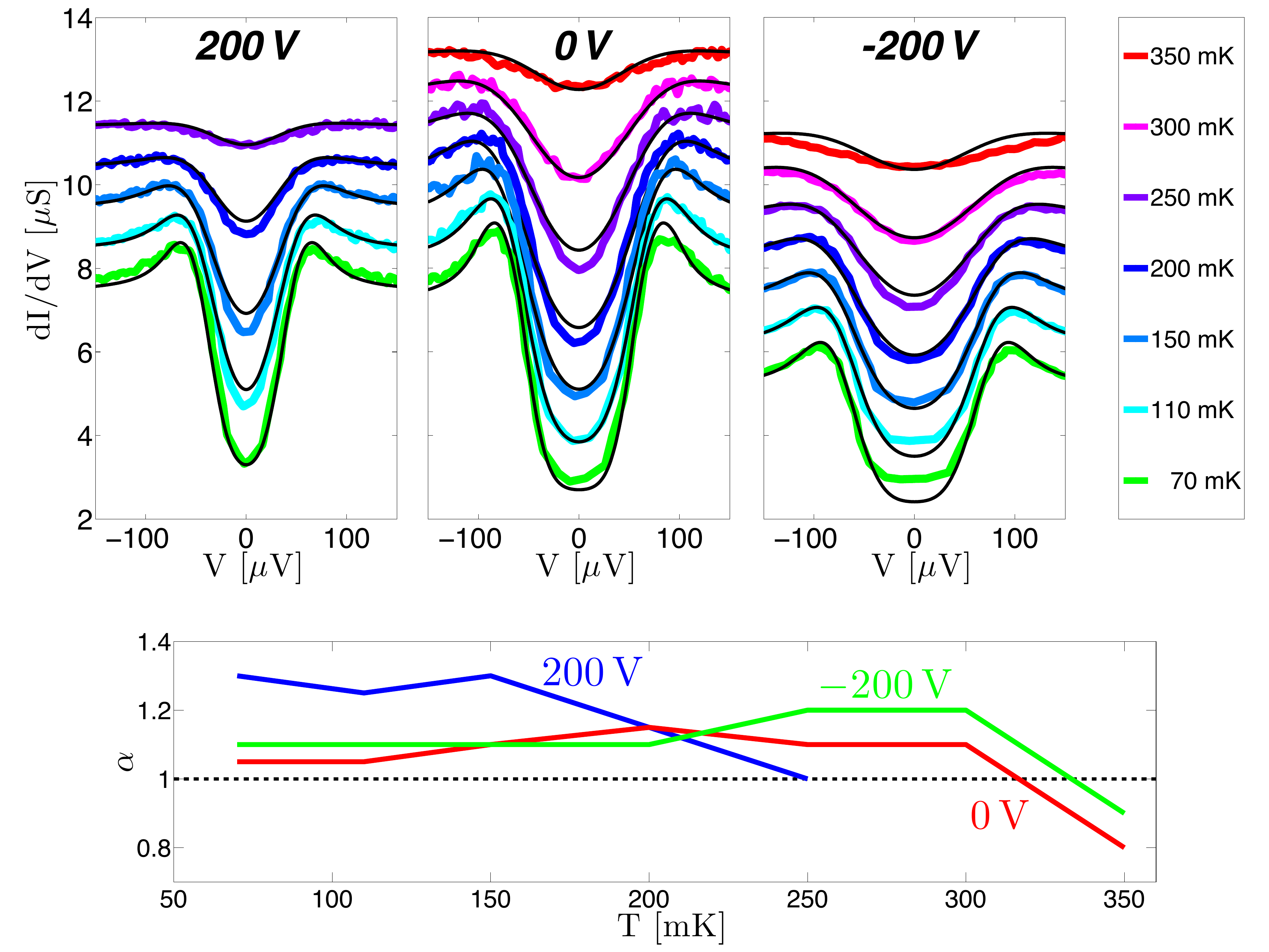}
\caption{Fits of experimental data of Ref. [\onlinecite{Richter}] with a Gaussian $P(T_c)$ and 
a variable coupling strength $\alpha(T)$. The parameters of the Gaussian fit are:
$V_G=200$\,V: $\overline{T}_c=244$\,mK, $\sigma=70$\,mK, $w=0.59$, $x=0.2$,
$V_G=0$\,V: $\overline{T}_c=383$\,mK, $\sigma=104$\,mK, $w=0.62$, $x=0.1$, 
$V_G=-200$\,V: $\overline{T}_c=383$\,mK, $\sigma=116$\,mK, $w=0.53$, $x=0.1$.
}
\label{SupplMat_alpha_di_T}
\end{figure}
We discuss the consequences of $\alpha(T)$ for the gating $V_G=200$\,V. There, the suppression in the DOS vanishes around $T^{\text{max}}_c\sim250$\,mK,
corresponding to a gap $\Delta^{\text{max}}\sim38\,\mu$eV (for $\alpha=1$). However, the low
temperature ($T=70$\,mK) spectra reveals gaps of at least $50\,\mu$eV and slightly smaller ones at intermediate temperatures ($T\sim150$\,mK). 
To obtain gaps of this order some islands with $T_c\sim340$\,mK have to be present in the system and, consequently, the fit of Fig. \ref{SupplMat_PTc_Vg200} is obtained with
parameters yielding non-negligible weight around this temperature ($\overline{T}_c=237$\,mK, $\sigma=151$\,mK, $w=0.67$, $x=0.1$). 
Clearly, such a distribution causes a too strong suppression for $T\sim250$\,mK.
Taking $\alpha(70\,mK)=1.3$ allows to take a smaller $P(T_c)$ shifted to lower temperatures ($\overline{T}_c=209$ and $\sigma=81$ $w=0.59$, $x=0.2$) thereby improving the fits at high temperatures
while still yielding good fits at low temperature as well. 
A similar reasoning applies to $V_G=0, -200$\,V. 
In addition there the high temperature curves $T=350$\,mK are best fitted with $\alpha<1$, signaling an attenuation of superconductivity.
This attenuation is reminiscent of the decrease in temperature of the coherent weight in Fig. 4 of the main text.

We conclude the considerations on the temperature dependence by looking at the effect of pair-breaking. 
Traditionally, this is done by means of a finite-lifetime-broadened density of states, the so-called Dynes formula:
\begin{equation}
\rho(E)=\text{Re}\Biggl[\frac{|E-i\Gamma|}{\sqrt{(E-i\Gamma)^2-\Delta^2}}\Biggr],
\end{equation}
where $\Gamma$ is a measure of the pair-breaking rate.
As before, we take a Gaussian $P(T_c)$ with $\overline{T}_c$, $\sigma$ and $w$ fixed in temperature (and $x=0$) such as 
to minimize the overall difference between the theoretical and the experimental curves for all temperatures
while $\Gamma(T)$ may vary with $T$. 
The discussion of results, reported in Fig. \ref{SupplMat_Dynes}, is somewhat delicate.
\begin{figure}[!htbp]
\centering
\includegraphics[scale=0.45]{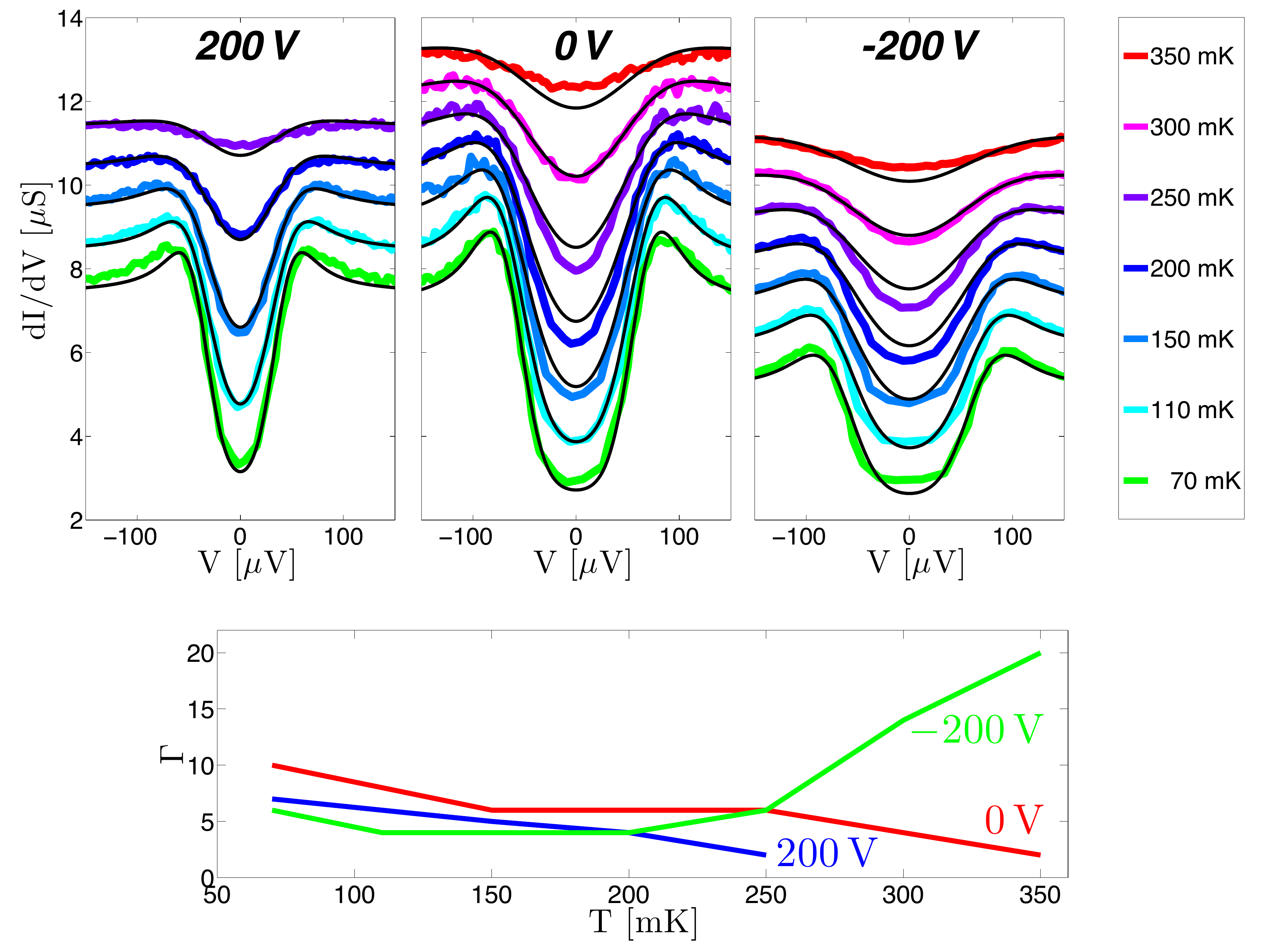}
\caption{Fits of experimental data of Ref. [\onlinecite{Richter}] with a Gaussian $P(T_c)$ a pair-breaking rate 
$\Gamma(T)$ (black solid lines).
The parameters of the Gaussian fit are:
$V_G=200$\,V: $\overline{T}_c=209$\,mK, $\sigma=81$\,mK, $w=0.6$
$V_G=0$\,V: $\overline{T}_c=383$\,mK, $\sigma=35$\,mK, $w=0.63$
$V_G=-200$\,V: $\overline{T}_c=418$\,mK, $\sigma=128$\,mK, $w=0.49$.
}
\label{SupplMat_Dynes}
\end{figure}
On general grounds one expects pair-breaking to be small at low temperatures, possibly 
becoming relevant at temperatures of the order of $T_c$.
While this is the case for $V_G=-200$\,V, the best fits for $V_G=200,0$\,V do not follow this expectation:
For these gatings, the parameter $\Gamma$ surprisingly decreases in temperature, while staying always smaller than $10\,\mu$eV.

In conclusion, the adjustments with $P(\Delta)$, $\alpha(T)$ and $\Gamma(T)$ do lead to an overall improvement of the fits in temperature.
Compared to the outcomes with $w(T)$ (Fig. 4 of the main text) however, the improvements are small. 
Nonetheless, the study of these various effects contributes to a better understanding of the temperature behavior of the DOS.
\clearpage
\section{Fitting Procedure}
The outcomes of this work rely on fits of the experimental data. In the following we explain in detail how these fits 
were obtained in the various cases.

{\bf T = 30\,mK, V$_G\in$ [-300,300]\,V:}
The data was extracted from Fig. 3(a) of Ref. [\onlinecite{Richter}]. 
The data was normalized by subtracting a straight line such that the $dI/dV$ values coincide
for $\pm250\,\mu$V and centered the data by a small horizontal shift (of the order of a few $\mu$V).
We discretized the curves in steps of $1\mu$V from $-250$ to $250\,\mu$V. For each set of parameters ${\overline{T}_c,\sigma,w,x}$ we calculated
the difference
\begin{equation}\label{diff}
D=\sum_{V=-250}^{250}\Bigg|\frac{dI}{dV}\bigl(V\bigr)\Big|_{\text{exp}}-\frac{dI}{dV}\bigl(V\bigr)\Big|_{\text{th}}\Bigg|,
\end{equation}
where `exp' and `th' denote the experimental and the theoretical value of the differential conductivity, respectively.
The best fit corresponds to the set ${\overline{T}_c,\sigma,w,x}$ which minimizes D.

{\bf T = 70\,mK, V$_G$ = 200, 0, -200\,V:}
We obtained the raw data for the behavior in temperature directly from the authors of Ref. [\onlinecite{Richter}].
We normalized the data by first fitting the normal state signal at $T=700$\,mK with a parabola, which we then subtracted
from the low temperature curves. Again, we centered the data by a small horizontal shift (of the order of a few $\mu$V).
We calculated the difference as in Eq. (\ref{diff}) for $V_G=200,0$\,V, while for $V_G=-200$\,V we enlarged the interval to $[-300,300]\,\mu$V.
As before, the best fit corresponds to the set ${\overline{T}_c,\sigma,w,x}$ which minimizes $D$.

{\bf T $\geq$ 70\,mK, V$_G$ = 200, 0, -200\,V:}
The analysis of $\alpha(T)$ and $\Gamma(T)$ were done by calculating $D$ for a set of parameters ${\overline{T}_c,\sigma,w,\alpha(T)}$ with $x>0$ fixed and ${\overline{T}_c,\sigma,w,\Gamma(T)}$ with $x=0$, respectively.
From all these sets we chose the one which minimized the sum of $D$ for all temperatures, with the constraint that ${\overline{T}_c,\sigma,w}$ be the same.
\end{widetext}
\end{appendix}

\begin{thebibliography}{99}
\bibitem{nota_pseudogap} To avoid any possible confusion about the notion of pseudo-gap state, we point out that 
in this work we intend by it a state which has non-zero resistance and yet exhibits a suppression of the DOS and 
more or less pronounced coherence peaks.
\bibitem{notaN0} The measured high-bias values actually correspond to an average of the DOS in the purely 
metallic regions, $N_0$, and of the DOS in the regions where pairing can occur, $\tilde{N}_0=N_0+\delta N_0$.
Then, the measured high-bias spectra, should correspond to a DOS $N_0+w_{pair}\delta N_0$. However, in the positive 
gating regimes $N_0\sim \tilde{N}_0$, while in the $V_G<0$ regime, where $\tilde{N}_0$ is substantially larger than
$N_0$, $w_{pair}$ rapidly decreases, again making rather inessential the difference between $N_0$
and the average measured value. Therefore, in order to keep minimal the number of fitting parameters, we
deliberately confuse this value with $N_0$ and fix this quantity from the measured high-bias spectra.
\end{thebibliography}
\end{document}